# Time Lag Evidence of Anti-Abortion Decree and Perturbation of Births' Distribution. A Benford Law Approach

**Author: Bogdan-Vasile Ileanu**

**Affiliation: (1) The Bucharest University of Economic Studies, (2) Center for Health Outcome & Evaluation Bucharest, Romania**

e-mail: bogdan.ileanu@csie.ase.ro; ileanub@yahoo.com

a. data availability statement

Part of the data that support the findings of this study are openly available at http://statistici.insse.ro:8077/tempo-online/, branch POP201B.

Part of the data that support the findings of this study are available from the corresponding author upon reasonable request.

b. funding statement**:**

This work was financed with the help of the projects:

*ReGrowEU / Advancing ground-breaking research in regional growth and development theories, through a resilience approach: towards a convergent, balanced and sustainable European Union / PN-III-P4-ID-PCCF-2016-0166*

*Understanding and modelling time-space patterns of psychology-related inequalities and polarization", code PN-III-P4-ID-PCCF-2016-0084*

**Aknowledgments:**

*Foremost, I thank Professor Marcel Ausloos, for all advice given in general related to research and publication. Also, I thank him for the valuable inputs given for the first draft of this paper..*

*Also, I thank Iulia Toma, who made an important effort and helped me to collect and verify some data in the old printed version of the Romanian Statistical Yearbooks.*

**Abstract**

This study analyzes the case of Romanian births, jointly distributed by age-groups of mother and father covering 1958-2019 under the potential influence of significant disruptors. Significant events such as anti-abortion laws application or abrogation, communism fall, and migration and their impact are analyzed. While in practice we may find pro and contra examples, a general controversy arises regarding whether births should or should not obey the Benford Law (BL). Moreover, the significant disruptors' impacts are not detailed discussed in such analysis.

I find the distribution of births is First Digit Benford Law (BL1) conformant on the entire sample, but mixed results regarding the BL obedience in the dynamic analysis and by main sub-periods. Even though many disruptors are analyzed, only the 1967 Anti-abortion Decree has a significant impact. I capture an average lag of 15 years between the event, the Anti-abortion Decree, and the start of distortion of the births' distribution. The distortion persists ~25 years, almost the entire fertility life (15 to 39) for the majority of a the people from the cohorts born in 1967-1968.

**Keywords:** disruptions, time-lag effects, abortion decree, births, Benford Law

**1 The general context**

First, the Benford Law is a quantitative technique successfully used in many fields to identify possible frauds or at least flag them. The existent studies do not evidence the existence of large time-lags between the possible disruptor and its effect(s), a fact which may conduct to some potential loss or lack link of actual causality between the real cause(s) and its or their effect(s).

Second, the problem of distribution of births and its obedience or non-obedience to BL doesn't have a solution yet, while pro and contra examples are present in the literature (Ausloos, Herteliu, and Ileanu 2015).

Third, the problem of pro-natality measures imposed by the Communist regime in many European countries after World War 2 (WW2) and later the Communism fall in the 1990s are extensively discussed (Girard 1983; Ross and Mauldin 1988), but some hidden effects remain.

And fourth, according to some authors (e.g., Flister 2013), these pro-natality measures, applied in Romania, were among the most drastic in the world. Besides the scientific attention, the media also puts enough light on the subject. For example, The Guardian (2013) underlines the importance of the issue too "On abortion, we should study Romanian history."

Thus, dealing with three intriguing and although challenging subjects, the current paper tries to bring evidence and answers on this demo-physics interdisciplinary topic, based on the Romanian data from 1960 to 2019.

For a better understanding of the foreword analysis, I organize the paper as follows: (1) A brief introduction on the Benford Law and its application; (2) The description of the significant disruptors and their impact on births data; (3) The data and methods used in the analysis. The study results can be read in section 4, and finally, in section 5, the reader may find the discussions and conclusions.

### 1.1 *The famous Law of Benford and its applications*

A general perception might be that the first digit (FD) of all numbers met in general, in nature, or in particular in a field of interest has an equiprobability of appearance, equal to 1/9. But, Newcomb (1881) and later Benford (1938) discovered that the numbers from nature have a curious property: Numbers starting with 1 are more frequent, the expected probability of their appearance is about 30%; meanwhile, the numbers starting with digit two are appearing in about 17% of cases, and so one, the numbers starting with 8 or 9 are found in less than 5% of the numbers coming from natural events. In fact, in Benford's (1938) paper, an excellent approximation of the probability of appearance of digit *d* on the first position is P(FD=*d*) = $log_{10}(1 + d^{-1})$, $d \in \{1…9\}$.

Later, extensions of BL1, such as the second digit law and the first two digits law, etc., are developed. Then, frequently, instead of studying only the distribution of the first digit, the second, the third digit distribution, and so on, are also analyzed. Frequently, in practice, the first two digits or first three digits combinations are studied. The extensions of BL1 law, i.e., BL2, BL12, BL3, BL123, etc., are derived by Hill (1995). Among other important considerations, Hill presents the "General significant digit Law," which gives the relation to computing the probability to encounter a digit d= {0, 1, …9} on any position 1 to…k, with the exception that 0 digits cannot be found on the first position.

Such a data property called Benford Law or "The Law of the Anomalous Numbers" or the First Digit Law (BL1) seems to guide a large number of natural facts, events, or processes like the lengths of the rivers, population, death rates (Benford 1938), distance to stars or galaxies (Alexopoulos 2014). This interesting law finds its applicability also in a wide area of domains. For example, Mebane (2006), using simulation procedures and BL test on Florida counties data from 2004, suggests that BL is an appropriate tool for auditing the election results. Pericchi and Torres (2011), using a Bayes approach, analyze the conformity of voting results in the USA, Venezuela, and Puerto Rico against BL1 and BL2 theoretical distribution. Mir (2012) finds BL conformity for adherents distributed by six of seven major religions. The author also remarks that cumulative followers of Christianity are not BL1 obedient. On the same topic, Ausloos et al. (2015) studies the daily number of births by religion, there orthodox vs. non-orthodox, from Romania and did not find any conformity to Benford Law. Moret et al. (2006) and Shukla et al. (2017) investigate different physical properties from BL perspective and confirm a good conformity of the law. The BL is a useful technique also in sociology. Diekmann (2012) proposes to apply the BL to increase the accuracy of sensitive questions. But, a huge momentum was found in economics. Here, the pioneer Nigrini uses the BL as an aid of detecting economic frauds. On the one hand, he proposed new methods to assess the BL obedience, such as MAD (Nigrini and Drake 2000) and, on the other hand, he and other colleagues successfully apply the BL properties in economic subfields to detect possible frauds (Nigrini 2015). Also, many other authors successfully used the BL to flag anomalies in the economic indicators (Mir Ausloos and Cerqueti 2014; Ausloos Cerqueti and Mir 2017; Shi Ausloos and Zhu 2018; Riccioni and Cerqueti 2018, etc.). Also, medical data are in the interest of BL analysis. For example, Crocetti and Randi (2016) deal with the cancer registries' data and find a very good BL1 conformity. Among other subjects, the current significant disruptor phenomena COVID-19 pandemic is investigated by many authors. For example, Sambridge and Jackson (2020) find that cumulative infections and the number of deaths from many countries are BL conformant. On the same topic, Lee, Han, and Jeong (2020) suggest that interventions, such as lock-downs or other restrictions, could cause the BL break-down even if the data source is not fake. In this case, the authors suggest that BL analysis of COVID-19 data would not be enough to characterize data as possibly fraudulent.

Briefly, the non-conformity to Benford Law is frequently associated with possible fraud or fake numbers or values or at least suspicious numbers, while a BL obedience is in general related to good, "natural data" or trustable data. But according to some authors, among others, Goodman (2016), the break-down of BL is not necessarily a fraud, it can be an anomaly, but it can have other plausible or natural explanations.

*1.2 The births in the local context*

The Second World War (WW2), which ends in 1945 has a considerable demographic impact on the entire world, first on the number of deaths and then on the number of births in the future periods.

Then, specific anti or pro natality legislative measures are adopted in many countries and also in Romania. First, an anti-abortion article was introduced in the Penal Code in 1948, and it is applicable until 1957. At the end of 1957, the Romanian Decrees 463 and 469 allow abortions. According to these laws, abortion becomes legal if it is approved by a medical commission and it became in the first three months of gestation (Massino 2019). Then a major event occurs in 1966, when the previous decrees 463 and 467 (Berelson 1976; LAW1 2021), are abrogated. Now, the 770/1966 decree forbids abortions, except under some particular conditions like illness of the mother, the mother is older than 45, the mother has already given birth to four children, and she takes care of them or in the case of incest or rape (Ghetau, 2018). In addition to these measures, the contraceptive import is stopped, and the marriage could be legally broken only in extreme circumstances. Since the beginning of the 1967 year, childless individuals or couples were forced to pay a tax, 10% or 20%, according to their income level. Besides the punitive measures, some benefits were also introduced, like supplementary income for women according to their number of children and child state allowance. (Moskoff 1980)

Various slight changes of the 770/1966 decree are added in 1972 and 1985. The Decree 53/1972, among other legal provisions, allows the age of the mother to be 40 instead of 45 to have a legal abortion. Again, in 1985 the previous 1972 legislation is changed. In Article 2/1985, women are allowed to abort if they are taking care of at least five children and also if they are older than 45 (Ghetau 2018).

As a result, according to World Bank data (World Bank, 2021), the fertility rate exploded from 1.9 births/ women in 1966 to 3.66 deliveries/ women in 1967. But, despite some additional adjustments of the existent law, such high values could not be maintained. Then, with a minor exception in 1974 and 1983-1986, a decreasing trend is registered. These draconic measures determine, according to Hord et al. (1991) and Kligman (1995), also the highest maternal mortality rates and among the highest infant mortality rates in Europe.

After the fall of Communism, on 26th December 1989 (Law2 2021), the 770 Decree was abrogated. Moreover, the 1990 year represents the starting year of the Romanian mass emigration flow. The period 1990-2019 is also under the influence of other major events. I note here that after Communism fall, the access of parents to education is more facile. Also, access to contraception methods significantly increases in the last decades.

Even if the access to information and access to contraceptive measures continue to grow, in Romania, the rate of abortions is one of the highest in Europe, reaching, in 2015, 359 procedures/1000 live births compared with 176 in Greece or 183 in Portugal in the same year (WHO 2021).

Despite possible larger access to contraceptives, access to knowledge, or the benefit of much more permissive laws, the downward trend of fertility also continues after 1990, and the fertility rate reaches its historical known minimum of 1.27 births per woman in 2002-03. After this minimum, the rate slightly increases, and in 2018 it reaches a value of 1.76 births/women, representing the maximum rate registered after the Communism fall.

From BL potential perturbators perspective, 1990-2019 is the period under the highest risk or the highest incertitude.

These significant events, spread over the 60 years, are very challenging for the perspective of possible natural obedience or failure of Benford Law. Thus, the impact of these events is uncovered in the results and discussion sections.

## 2 Data and methods

*2.1 Data sources and data structure*

I collected the data related to the joint distribution of the number of births by mother and father age groups for the period 1958-86 and 1989-2019.

The entire data collected for the period 1958-85 and 1988-89 was input from Statistical Yearbooks found in the Bucharest University of Economics Digital Library, National Institute of Statistics

Library, and personal library. All the data corresponding to the period 1990-2019 was automatically extracted in *.csv files from the National Institute of Statistics, On-line Tempo Database (Tempo, 2021). Due to the Communism policy of austerity, the volumes of Statistical Yearbooks from 1987-89 were significantly reduced, and the joint distribution of the number of births by age-group of mother and father was not published. I found no data for these years in other public sources.

Thus, for each year t=1958-1986 ∪ 1989-2019, I defined the following matrix of data:

n (i, j) = number of live births conditioned by *i* and *j*, each of them ∈ {1...10}, the 5th age-group of the mother. The age groups are: {<15 y.o.; 15-19; 20-24; 25-29; 30-34; 35-39; 40-44; 45-49; >49 and unknown/non-declared}. In this case, a theoretical three-dimensional data table is available with a total of 60x10x10=6,000 cells. The cells with the number of births declared with both age-group of mother and father undeclared are scarce, counting only 36. An example of this multi-dimensional table used in the analysis for specific years is given in Table 1.

| Year (t) | Mother's age-group | <15 y.o. | 15-19 | 20-24 | 25-29 | 30-34 | 35-39 | 40-44 | 45-49 | 50 and over | Unknown |
|---|---|---|---|---|---|---|---|---|---|---|---|
| | | | | | Father's age-group | | | | | | |
| 1958 | <15 y.o. | - | 25 | 46 | 34 | 9 | 4 | 1 | - | - | 43 |
| 1958 | 15-19 | - | 5,649 | 15,740 | 17,148 | 1878 | 202 | 28 | 17 | 11 | 20 |
| 1958 | 20-24 | - | 1,790 | 43,747 | 71,055 | 20,061 | 2,313 | 373 | 178 | 84 | 61 |
| 1958 | 25-29 | - | 242 | 8,508 | 50,284 | 37,783 | 9,817 | 1,675 | 737 | 302 | 50 |
| 1958 | 30-34 | - | 42 | 1,080 | 8,742 | 26,550 | 17,013 | 5,144 | 2,305 | 987 | 32 |
| 1958 | 35-39 | - | 12 | 147 | 1,097 | 4,355 | 11,000 | 7,212 | 4,885 | 1,797 | 40 |
| 1958 | 40-44 | - | 7 | 10 | 55 | 231 | 595 | 1,750 | 2,556 | 1,244 | 11 |
| 1958 | 45-49 | - | - | - | 7 | 11 | 35 | 117 | 500 | 451 | 4 |
| 1958 | 50 and over | - | 2 | 4 | 7 | 6 | 2 | 2 | 7 | 36 | 5 |
| 1958 | Unknown | - | 2 | 81 | 75 | 53 | 24 | 10 | 9 | 7 | 279 |

| Year (t) | Mother's age-group | <15 y.o. | 15-19 | 20-24 | 25-29 | 30-34 | 35-39 | 40-44 | 45-49 | 50 and over | Unknown |
|---|---|---|---|---|---|---|---|---|---|---|---|
| 2019 | <15 y.o. | 3 | 190 | 134 | 28 | 10 | 4 | 1 | 1 | 1 | 328 |
| 2019 | 15-19 | 3 | 2,233 | 6545 | 3,068 | 1,178 | 290 | 94 | 19 | 18 | 3,383 |
| 2019 | 20-24 | - | 347 | 7255 | 14,039 | 7,208 | 1,633 | 426 | 131 | 53 | 2,509 |
| 2019 | 25-29 | - | 46 | 1447 | 18,228 | 22,866 | 6,538 | 1,573 | 297 | 103 | 1,686 |
| 2019 | 30-34 | - | 18 | 324 | 4,174 | 23,875 | 16,501 | 4,995 | 1,002 | 306 | 1,353 |
| 2019 | 35-39 | - | 5 | 61 | 611 | 3,261 | 10,044 | 6,508 | 1,478 | 478 | 798 |
| 2019 | 40-44 | - | 2 | 17 | 79 | 377 | 1,111 | 2,354 | 1,036 | 391 | 274 |
| 2019 | 45-49 | - | - | - | 3 | 15 | 26 | 61 | 106 | 75 | 23 |
| 2019 | 50 and over | - | - | - | - | 3 | 1 | 2 | 3 | 6 | 4 |

Table 1 The distribution of births by age mother and father in the first and last available year

*2.2 Methods of analysis*

The Benford law analysis is made using the following approaches.

(a) First, I studied the distribution of births by age group of both parents for the entire period 1958 to 2018. The analysis takes into account all the non-null cells from a total of 6,000.

(b) Second, I split the entire time interval into three sub-periods, P1, P2, and P3, taking into account the major events described in the Introduction:

The first sub-period, P1, lies between 1958 to 1966. Here, the analysis should capture the influence of the 1957 Decree and some effects of WW2.

P2 - covers the period from 1967 to 1989. Here, the results are expected to be dominated by the 770/1966-abortion decree and later additional changes.

P3 - follows the period from 1990 to 2019. The latest period should capture multiple actions such as abrogation of the 770/1966 decree, migration, but also more extensive access of the population to higher education.

(c) In the third stage, I made a time-dependent analysis, computing relevant indicators for testing BL1 every year since 1966. This approach underlines the dynamic conformity to BL1 in a more detailed manner.

A lot of challenges regarding the methods used to test the BL conformity can be found in the current literature, for example, in Lesperance et al. (2016) or in Joannes-Boyau and colleagues (2015). Taking into account the entire sample and sub-samples dimensions combined with the conclusion of Lesperance and colleagues (2016), regarding its power against other tests, the classical Chi-squared ($\chi^2$) test is an appropriate option for the present analysis.

The statistics used for the analysis are: (1) the static classical $\chi^2 = \sum_{i=1}^{9} \frac{(Oi - Ei)^2}{Ei}$ where $Oi$ = the observed counts for digit $i$ and $Ei$ = the expected counts for the digit $i$ according to BL1. This statistic is applied for the entire sample, for each subperiod P1 to P3, but also for the time-dependent analysis. When $\chi^2$ statistics are computed, in the time-dependent analysis, I counted all the values for each first digit, from 1958 to the moment τ, for any τ>=1966. In this case, the relation becomes $\chi^2(\tau) = \sum_{t=1958}^{\tau} \sum_{i=1}^{9} \frac{(Oit - Eit)^2}{Eit}$, where τ, counts for all years in the closed interval 1966-2019. Here, the critical value for type I error of 0.05 and 8 degrees of freedom, equal to 15.50, can be found in any introductory statistics textbook.

NB: While the data are missing for the years 1987 and 1988, it is easy to understand that $\chi^2(1986) = \chi^2(1987) = \chi^2(1988)$. The usage of $\chi^2(\tau)$ may be criticized when comparing the values in dynamics since this measure is strongly dependent on the sample volume. Then, I used the measure $\chi^2(\tau)/n$, and I compared the values with the critical $\chi^2(0.05,8)=15.5/n$ to underline the non-significant effects of the sample volume in the analysis.

One may criticize the choice of Chi-squared statistics and the post-comparison with critical $\chi^2(\tau)/n$, mainly because the results are strictly dependent on sample volume (n). Thus, I also use the mean absolute deviation (MAD) measure proposed by Drake and Nigrini (2000). The MAD is not a statistical test; it rather shows how close or how far we are from BL conformity. According to the authors, if MAD is below 0.004, then we have close agreement, a MAD between 0.004 and 0.008 shows acceptable conformity, finally, a MAD over 0.012 flags a non-conformity to BL1.

### 3  Results

3.1 General aspects

First, the natality rate has a high fluctuation in the analyzed period. Its dynamic is mainly the result of the socio-economic changes, but in the same measure, it is the result of the multiple legislation changes in this long period.

The number of births grouped simultaneously by the age of mother and father has a very large variation between the groups in a given year. In general, the lowest non-zero values are registered in those cases when parents have extreme age groups. For example, in 2019, in Romania, there was one birth for age mother under 15, and age of the father was between 45 and 49, and one birth when the mother had over 50 years and the father was in the 35-39 age group. But most births occur between 20-35 years old of both parents. For such a long period, analyzed here, characterized by so many socio-economic and political events, it is normal to see significant changes during time. For the curiosity of the reader, I present in Figure 1 the structure of births by major age groups of parents.

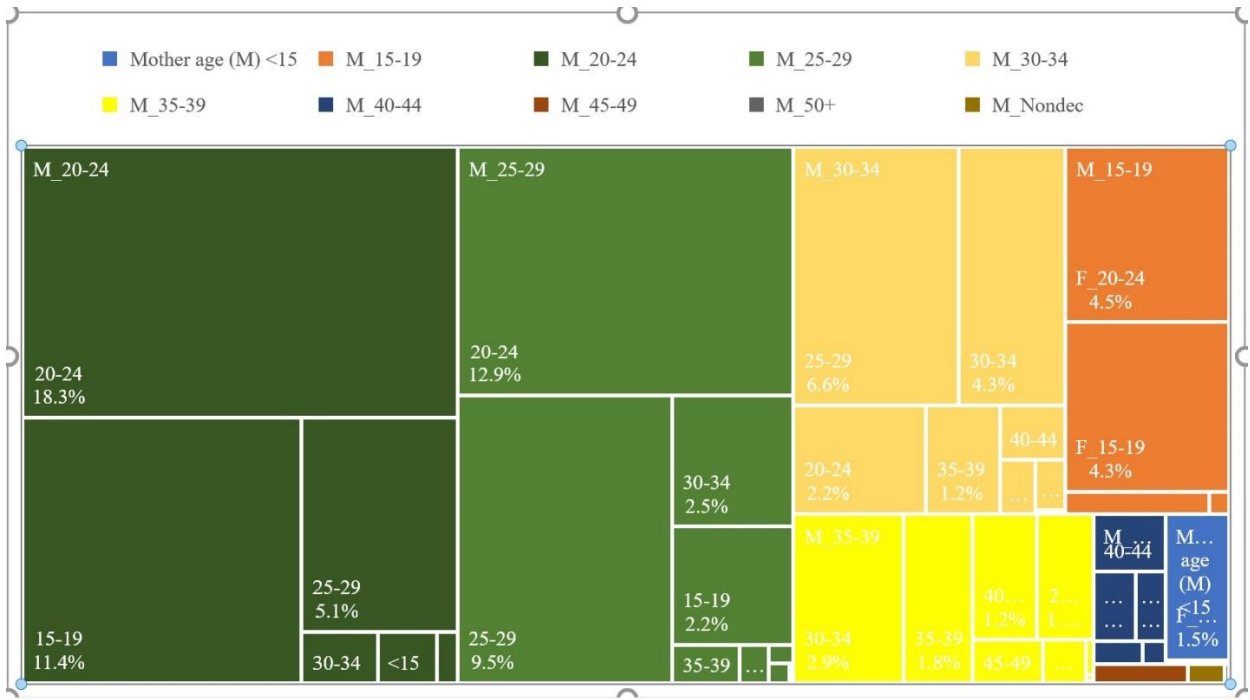

Figure 1. The structure of births by the age of mother and father, accumulated data for 1958-59

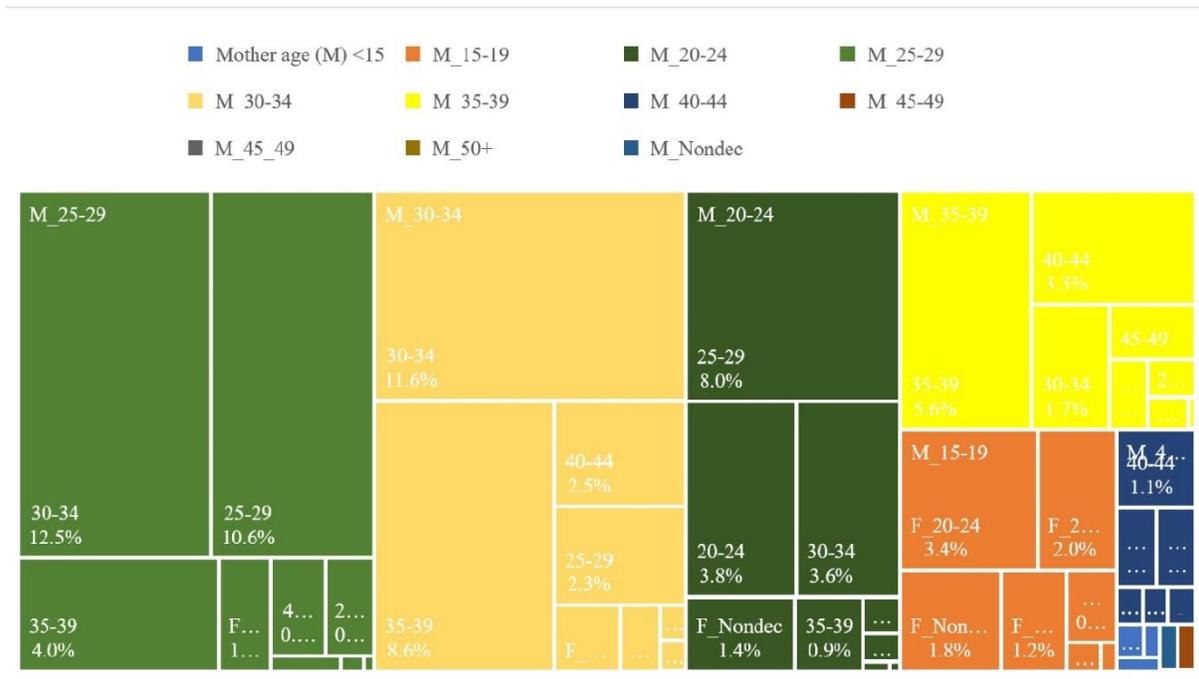

Figure 2. The structure of births by the age of mother and father, accumulated data for 2018-19

The changes occurring during the six decades are shinned in Figures 1 and 2. In 1958-59, the leader combination of age-groups for mother and father were the same 20-24-year-old. Then, almost 20% of total births from a year occurred in that combination of age groups. In the second place, as a share of births, the combination between a mother aged 25-29 and a father aged 20-24 had an important contribution, 13% of total yearly births being in this category. The extreme age groups were rarely found. Sixty years later, in 2018-19, the situation is visibly different. The top category is given by the combination of a mother aged 25-29 and a father aged 30-34 years old. This result shows an increase, for the age of birth, in mean, for five years in mother's case and ten years for the father. Moreover, the second combination is the frequency of births in the 30-34 years old for both parents. One may observe that extreme groups such as 40-44 years for both parents become visible in the context of share of births. The structure of data, based mainly on 9 x 8 x 59 cells with data combined with a high variation between age groups and during the time, is an interesting candidate for a Benford Law analysis.

3.2 Anomalies detection

On the overall sample, I observe tight conformity to Benford Law. Even if there are doubts regarding the usage of the Chi-Squared test in the cases when the sample is large, in our case, the computed value of this statistics is 9.52, much lower than the correspondent 0.05 critical value, here, for 8 degrees of freedom, 15.50.

Broadly speaking, the results achieved on the overall sample does confirm the hypothesis that the number of births is BL conformant. Here, even if they are split by age group of mother and father, they remain conformant. Due to the fact that the data cover here an extensive period, with many socio-economic changes, I break the entire sample into sub-samples as described in the second sub-section of this paragraph.

| First Digit | BL1 expected (%) | Empirical BL1 frequencies (%) | | | |
|---|---|---|---|---|---|
| | | Total (1958-2019) | Period sub-samples | | |
| | | | 1958-66 | 1967-86+1989 | 1990-2019 |
| 1 | 30.1 | 31.3 | 35.4 | 32.7 | 29.3 |
| 2 | 17.6 | 17.2 | 16.8 | 16.1 | 18.0 |
| 3 | 12.4 | 12.4 | 12.5 | 11.7 | 13.0 |
| 4 | 9.6 | 10.4 | 9.9 | 10.9 | 10.2 |
| 5 | 7.9 | 7.8 | 6.1 | 8.0 | 8.2 |
| 6 | 6.6 | 6.0 | 3.8 | 6.6 | 6.5 |
| 7 | 5.7 | 5.9 | 7.8 | 4.7 | 6.0 |
| 8 | 5.1 | 4.9 | 4.5 | 4.8 | 4.7 |
| 9 | 4.5 | 4.2 | 3.1 | 4.6 | 4.1 |
| Sample (n) | - | 4 243 | 704 | 1 414 | 2 125 |
| Computed $\chi^2$ | - | 10.56 | 26.94 | 11.31 | 3.68 |

Table 2 The observed and expected frequencies in the case of BL1 by different sub-periods

When the first-time window is analyzed, 1958-1966, the BL1 is broken down, while the computed Chi-squared exceeds the critical value at 0.05 level. The reasons why the BL1 is not obeyed are various. For example, I may claim that this result is a lagged effect of decree no. 463, which allowed abortion. Abortion availability may have a significant impact on the natural course of births. Moreover, I also may claim the poor quality of the birth classification by the age of mother and father. The post-WW2 period was challenging from any socio-economic perspective, thus in particular for the data collection. However, the empirical Chi-squared of 19.27 is not very far from the 0.05 critical value of 15.50, showing that even if some abnormal behavior persists, it is not very powerful, or its strength decreased in time until the point of measurement.

In the second time window, 1967 to 1986 ∪ 1989, I expect to see the results of the 1966 abortion decree and extra restrictive measures. But the product is strange in the sense of BL conformance. When the computed Chi-squared has a value of 12.77, I cannot reject the hypothesis of the BL1 conformance assuming a 0.05 type I error. At this point, I instead conclude that 770-Decree has no significant impact on the BL conformance.

In the third time window, covering the 1990-2019 period, I expect to measure the impact of the migration flow started after 1990 and the impact of the abrogation of the 770 Decree. When the results are analyzed, one can see that BL1 is conformant. Moreover, the computed Chi-squared has an unexpected low value-3.68.

The question remains how is it possible that such dramatic events have no impact on the BL1 conformance. It is possible for the analysis by sub-periods/time window, made in this classical way, to be not relevant enough. As a result, I computed the time-dependent Chi-squared, $\chi^2(\tau)$, from the starting period to the moment $\tau$. All the values are in Figure 3.

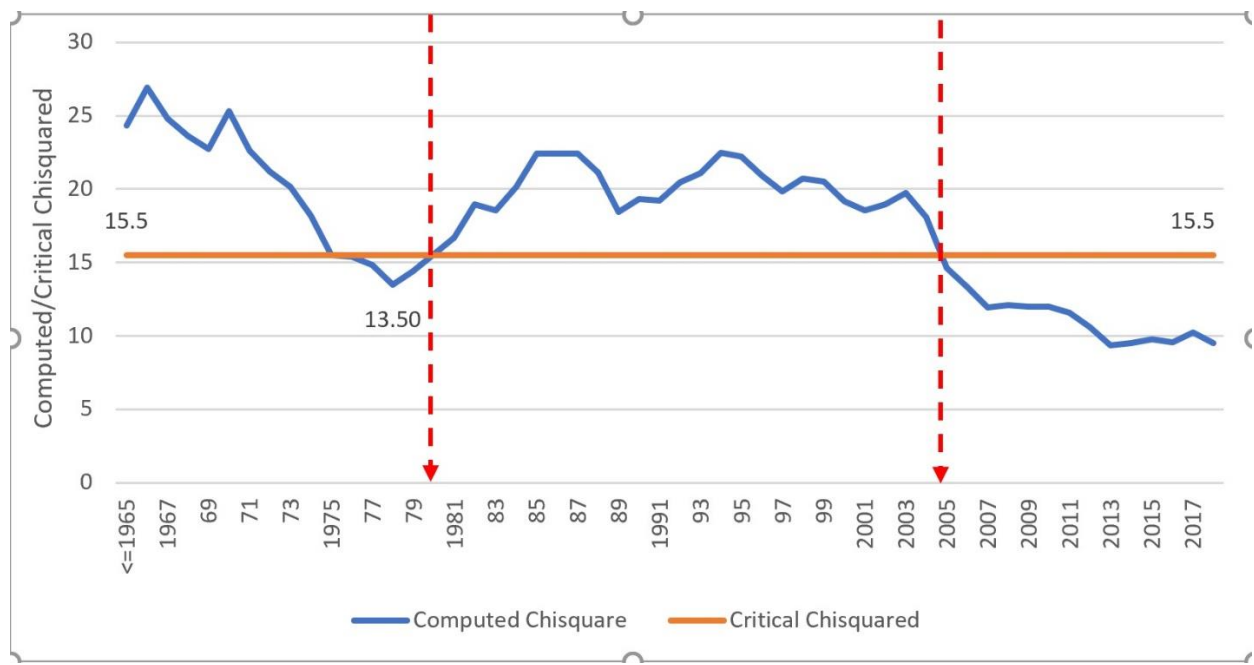

Figure 3. The dynamic of Computed Chi-Squared vs. 95% Critical Chi-squared divided by sample size, during 1965-2019

The Chi-squared has a sinuous evolution showing a break-down of BL1 conformity in the period 1981-2005. After 2005, the empirical value has a visible decreasing trend to the end of the time series.

One can claim that the dynamic analysis of Chi-squared maybe not be robust because, theoretically, the computed Chi-squared is an increasing function of sample volume. Then, in Figure 4, I present the dynamic of empirical $\chi^2(\tau)$ divided by sample volume.

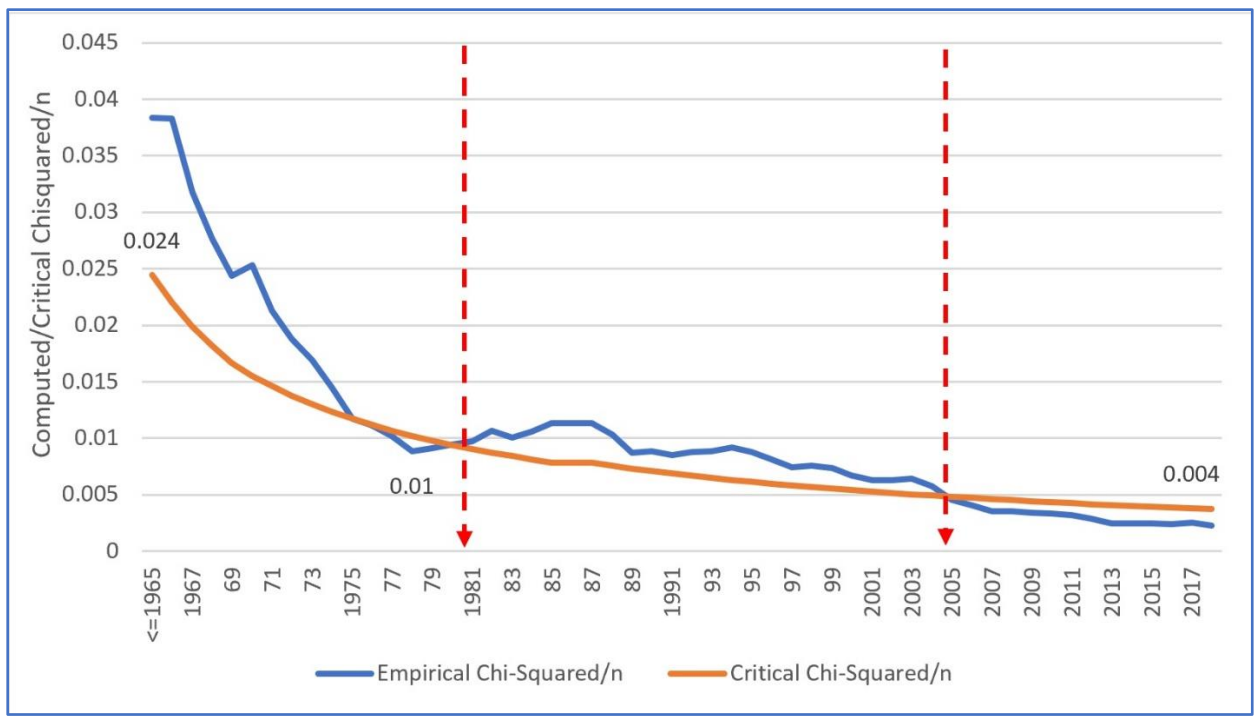

Figure 4. The dynamic of Computed Chi-Squared vs. 95% Critical Chi-squared divided by sample size, during 1965-2019

The evolution of the empirical value divided by sample volume keeps the behavior of the initial value. Even if the amplitude of deviations to the critical curve is lower, the conclusion is the same. There is one period of BL1 break-down, before 1975, which may be attributable to WW2, and then a large period lying between 1981 to 2005, which is related to the 1967 Abortion Decree. After 2005, the 1966 Decree effects disappear, and the distribution of births returns to its BL1 natural conformity. In Figure 5, I plot the MAD values and the Nigrini reference values.

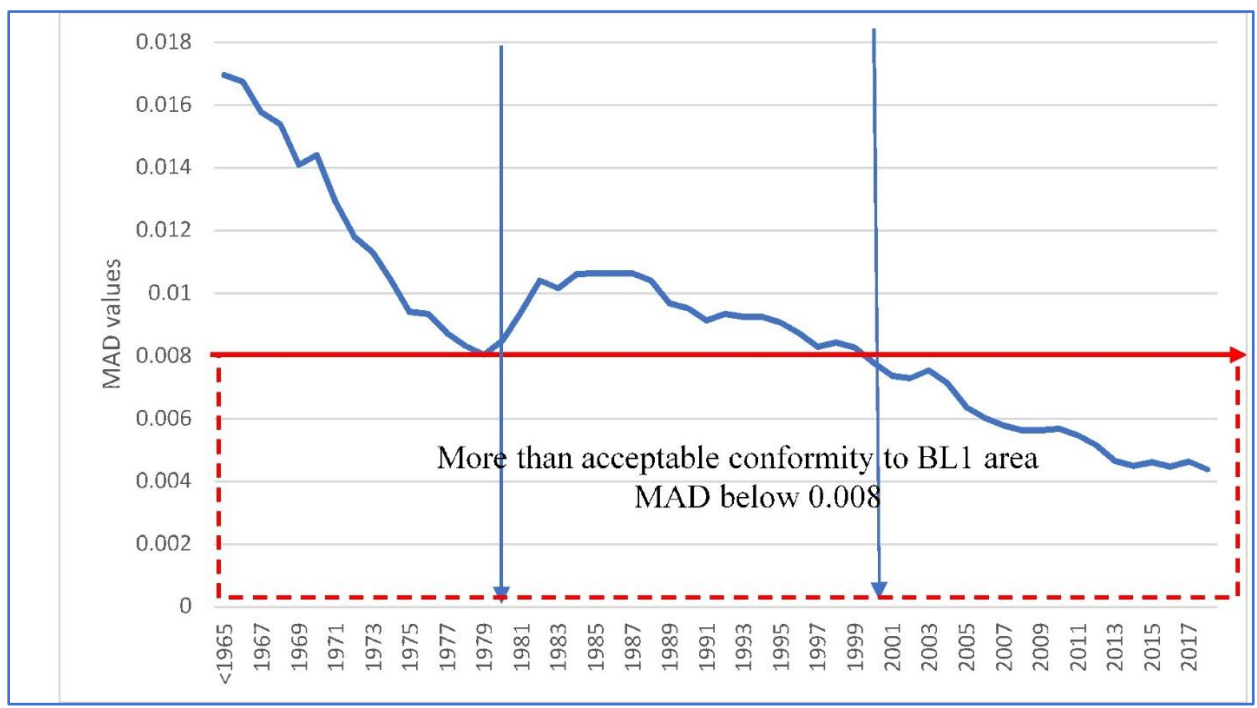

Figure 5. The dynamic of MAD values and area of BL1 acceptable conformity

Almost like in the case of Chi-squared statistics, the MAD values dynamic follow the same behavior, sustaining the break-down connection with the previously significant events, here WW2 and 1966 Decree.

## 4  Discussion and conclusions

Considering that the number of births is one of the essential socio-economic indicators in every country, I expect registration and reporting errors to be insignificant. In this case, one should not expect a non-obedience to BL, while there are multiple examples where the number of births is BL conformant. The results achieved here are mixed. First, I found that on the entire sample, the number of births is BL1 conformant. But this result may represent a trap. The dilemma regarding this possible trap is supported by the fact that it is almost impossible that major events such as the Second World War, three legislative measures against or pro natality, and the open gate of the migration have no significant effect when testing the BL conformance. The BL1 tests applied by sub-samples flag an anomaly for the 1958-66 period, somewhat attributable to WW2 effects and the 1957 decree. Unfortunately, the BL does not detect the impact of the other events when it is applied on the sub-samples 1967-89 and 1990-2019. When the dynamic analysis is made, a clear breakdown of BL1 is visible from 1982 to 2003-2005.  Considering that between 1980 and 1989, there are no radical policies pro or against natality, then the BL1 break-down may be attributable to the 1966 abortion decree combined with the other restrictive measures applied since 1$^{st}$ January 1967. Thus the 1982 to 2003-2005 period is a 15 to 39 years lag distance related to that critical event.

Moreover, as it is mentioned earlier in this paper ~80 % of total births in a year, are between 15-39 years - the maximum fertility age interval of both mother and father. The BL break-down covers the entire reproductive life period of the 1967 and 1968 cohorts. The impact of sample volume is not a deterministic cause of the BL break-down since at the end of the time series, after 2005, when the sample volume significantly increases, reaching more than 4200 units at the end of the analyzed period, the computed Chi-squared dramatically drops down. The Nigrini MAD value, independent of sample value, confirms the results achieved by Chi-squared statistics.

In the case of time series, the usage of the BL to flag anomalies, particularly in the demographic phenomenon, should be applied with caution when between disruptor and effect exists a lag. Here, when the disruptors were acting in 1966, BL detects anomalies between 1983 and 2006. The lag period depends on the analyzed phenomenon, while the magnitude and the length of the disturbance detected by BL are dependent on multiple factors. Here, the perturbation length is strongly associated with the time passed by adults during the productive life, from 15 to 40, approximatively 25 years.

As other authors noticed (Druica et al. 2018), one variable or one phenomenon should not receive a verdict such as BL conformant or not obedient. One phenomenon could have multiple states during a given period. Studying the results published in Ausloos et al. (2015) regarding the daily number of births during the 97 years analyzed linked with the results from this paper regarding the number of births jointly distributed by age-group of mother and father, broadly speaking, we have different effects. As a result, it is possible that one phenomenon, analyzed (even) in the same period, but grouped from a different perspective, may have both states: BL conformant but also BL not obedient.  These results may also impact other demographic phenomena such as the number of deaths/mortalities, population, drugs and therapies effects, or other dynamic series where lags between action and results are present. In the current context of COVID-19 pandemics, according to these results, the impact of COVID-19 pandemics as a particular disruptor may be seen on demographics phenomena now, but even after 15-20 years…

Another point that should be noticed is that the analyzed series tends to have oscillations related to its departure to BL1 conformity. Then, in the long term, this could be a property of a very long time series. The current set of data is not long enough to justify the existence of cyclicity or other periodical departures or closures to BL1 conformity, but the idea can be explored in future studies.

27. Pericchi, Luis and David Torres. 2011. Quick Anomaly Detection by the Newcomb–Benford Law, with Applications to Electoral Processes Data from the USA, Puerto Rico, and Venezuela. Statistical Science 26: 502-516. http://dx.doi.org/10.1214/09-STS296

28. Riccioni, Jessica and Roy Cerqueti. 2018. Regular paths in financial markets: Investigating the Benford's law. Chaos, Solitons and Fractals 107: 186-194, https://doi.org/10.1016/j.chaos.2018.01.008

29. Ross John A. and Parker W. Mauldin. 1988. Romania's 1966 Anti-Abortion Decree: The Demographic Experience of the First Decade. In: Ross J.A., Mauldin W.P. (eds) Berelson on Population. Springer, New York, NY. https://doi.org/10.1007/978-1-4612-3868-3_5

30. Sambridge, Malcolm and Andrew Jackson. 2020. National COVID numbers — Benford's law looks for errors, Nature, 384, https://dx.doi.org/10.1038/d41586-020-01565-5

31. Shi, Jing, Marcel Ausloos and Tingting Zhu. 2018 Benford's law first significant digit and distribution distances for testing the reliability of financial reports in developing countries, Physica A. 492: 878-888. http://doi.org/10.1016/j.physa.2017.11.017

32. Tempo (2021) http://statistici.insse.ro:8077/tempo-online/#/pages/tables/insse-table/Root Social Statistics, Branch-A2/Natality/POP201I. Accessed on 24th April 2021

33. The Guardian (2013) https://www.theguardian.com/commentisfree/2013/jan/15/abortion-romanian-history Accessed on 24th April 2021.

34. Law1 http://legislatie.just.ro/Public/DetaliiDocumentAfis/177

35. Law2 http://www.cdep.ro/pls/legis/legis_pck.htp_act_text?idt=11000

36. World Bank (2021a) https://data.worldbank.org/indicator/SP.DYN.TFRT.IN?locations=RO, Accessed on 24th April 2021.

37. WHO (2021) https://gateway.euro.who.int/en/indicators/hfa_586-7010-abortions-per-1000-live-births/ Accessed on 24th April 2021.